# Physical Mechanism of Nuclear Reactions at Low Energies

V.P. Oleinik* and Yu.D. Arepjev**

*Department of General and Theoretical Physics,
National Technical University of Ukraine **"Kiev Polytechnic Institute"**,
Prospect Pobedy 37, Kiev, 03056, Ukraine
**Institute of Semiconductor Physics, National Academy of Sciences,
Prospect Nauky 45, Kiev, 03028, Ukraine; e-mail: yuri@arepjev.relc.com
http://www.chronos.msu.ru/lab-kaf/Oleynik/eoleynik.html.

**Abstract.** The physical mechanism of nuclear reactions at low energies caused by spatial extension of electron is considered. Nuclear reactions of this type represent intra-electronic processes, more precisely, the processes occurring inside the area of basic localization of electron. Distinctive characteristics of these processes are defined by interaction of the own field produced by electrically charged matter of electron with free nuclei. Heavy nucleus, appearing inside the area of basic localization of electron, is inevitably deformed because of interaction of protons with the adjoining layers of electronic cloud, which may cause nuclear fission. If there occur "inside" electron two or greater number of light nuclei, an attractive force appears between the nuclei which may result in the fusion of nuclei. The intra-electronic mechanism of nuclear reactions is of a universal character. For its realization it is necessary to have merely a sufficiently intensive stream of free electrons, i.e. heavy electric current, and as long as sufficiently great number of free nuclei. This mechanism may operate only at small energies of translational motion of the centers of mass of nuclei and electron. Because of the existence of simple mechanism of nuclear reactions at low energies, nuclear reactor turns out to be an atomic delayed-action bomb which may blow up by virtue of casual reasons, as it has taken place, apparently, in Chernobyl. The use of cold nuclear reactions for production of energy will provide mankind with cheap, practically inexhaustible, and non-polluting energy sources.

## 1. Introduction

> Tell me what the electron is, and I shall
> explain to you everything else.
> W. Thomson

Nuclear reactions at low energies, occurring in physical and biological systems, and, in particular, the cold fusion (CF) of nuclei, attract ever increasing attention (see review articles [1,2]). This is explained by the fact that research on CF (in what follows, by cold fusion we shall understand any nuclear reactions at low energies) opens up the way to the solution of the problem which was set more than 50 years ago in the field of controlled thermonuclear reactions (CTR) and which has not been solved - to provide mankind with cheap fuel. An important point is that CF allows to create not only cheap, but also non-polluting energy sources, as nuclear reactions at low energies are not accompanied by radiations dangerous to health ($\gamma$-radiations, streams of fast neutrons and other particles). Note that the energetic problem facing mankind is presently of special interest in connection with the fact that, according to expert evaluations, the oil-and-gas resources in the world will suffice only for some decades. For this reason the study of CF is among the most important problems of physics.

It is necessary to note that, relying on the standard theory of nuclear reactions describing nuclear processes in vacuum, experts in the field of nuclear physics, engaged in CTR, reject the very possibility of existence of nuclear fusion at low energies. Two basic objections are raised against CF:
1. at low energies the penetrability of Coulomb barrier around nuclei is so small that the probability of nuclear fusion is practically equal to zero;
2. distinction between the atomic and nuclear energy scales is so great that the energy, which might be evolved as a result of nuclear fusion, could not be transferred directly to atomic lattice; therefore the energy above should be emitted in the form of streams of $\gamma$-quanta, fast neutrons and other particles. However, such streams of sufficient intensity have not been registered.

The answer to the first objection against existence of CF is that at the heart of CF are nuclear processes occurring in environment, and the basic role is played here, apparently, by collective effects caused by interaction of nuclei with particles of environment in which the nuclear reaction takes place. The laws governing the behaviour of interacting nuclei in vacuum are inapplicable to the description of CF of nuclei [3]. Nuclear reactions occurring at low energies submit to completely different laws which can be established only provided that collective effects mentioned above are taken into account. For this reason the standard theory of nuclear reactions in vacuum can by no means refute the existence of CF.

As to the impossibility of transferring the energy between levels of various scales, we can give an example of the phenomenon of sonoluminescence (luminescence of a liquid when a sound wave causing cavitation passes through it) [4], in which the energy transfer from an acoustic wave to electromagnetic field occurs with appreciable probability in spite of the fact that the distinction between energies of acoustic phonons and quanta of light reaches 11 orders.

As early as 10 years ago J. Schwinger, the Nobel winner and the known expert in the field of the theory of elementary particles and quantum electrodynamics, asserted that it is impossible to deny the reality of CF phenomenon [3,4]. Since then the CF phenomenon for nuclei was repeated hundreds times in laboratories all over the world, tens of patents on the ways of energy generation on the basis of CF were registered and enormous number of experimental works were published, which not only confirm the existence of effect, but also contain its detailed analysis.

The most convincing evidence for the existence of nuclear reactions at low energies seems to give the mass-spectrometric research of reaction products [5] as well as research on biological systems [6]. Detailed study of electric explosion of foil made of especially pure materials in water, described in [5], suggests that at electric discharges transformation of chemical elements occurs. Study of optical spectrum of plasma arising at discharge and of the mass-spectrometric analysis of sediments, which remained after the discharge, shows that there appears in plasma a significant number of chemical elements which were not present in the initial material of explosive foil and electrodes and also that the isotope structure of the foil material changes appreciably. The change of experimental conditions, for example, of energy contribution in foil, its mass and dimensions results only in redistribution of intensity of plasma spectral lines, i.e. in the change of statistical weight of chemical elements in plasma, but the composition of chemical elements remains unchanged and it essentially depends on the material of foil. As is seen from the results received, nuclear reactions taking place at electric discharge are not accompanied by the occurrence of a stream of neutrons and $\gamma$-radiation and proceed at low energies of atomic nuclei.

**The research mentioned above** as well as many others, carried out by different researchers in different laboratories, **allow to draw a conclusion that existence of nuclear reactions at low energies is reliably established.**

The development of research on CF is hampered by the absence of theory of the phenomenon. As is noted by Schwinger [3,4], the situation in CF closely parallels that in high-temperature superconductivity: reality of the last, as a result of careful experimental research, is completely established, though theory of the phenomenon is absent till now.

In [5], to account for the transformation of chemical elements, the hypothesis is put forward that at the electric explosion of foil in the plasma channel are formed magnetic monopoles which may overcome the Coulomb barrier even at insignificant kinetic energy due to the great magnitude of their magnetic charge. The monopole, appearing not far from a nucleus, causes its polarization: those nucleons of the nucleus, which are situated more close to the monopole, experience stronger influence of the last, than the nucleons situated on the opposite side of the nucleus. As a result, a deformation of the nucleus arises (the nucleus is lengthened), which may result in nuclear fission.

Obvious drawback of this mechanism of nuclear reactions is that magnetic monopoles have yet to be found out in nature.

Numerous attempts to construct a consistent theory of CF (see reviews [1,2]) have not been crowned with success. As it was noted above, for the CF to be described, the account of the collective effects may be important caused by interaction of nuclei with environment, in which nuclear reaction takes place. But does it suffice to take into account these effects in order that the theory of the phenomenon be constructed? The analysis of the experiments on transformation of chemical elements at low energies and on the CF of nuclei suggests that the phenomenon discussed does not fall within the domains of exotic ones: it seems to occur in nature constantly, at every step, in both physical and biological systems. Therefore, it is natural to expect that nuclear reactions at low energies should have a simple physical explanation.

However such explanation, which is not beyond the scope of existing representations, is yet to be found. Doesn't it mean that we are facing here the situation similar to that which has arisen in physics at the end of the 19th century and which has been figuratively described in the words: on the light sky of physics there are only two small dark clouds – the radiation of absolutely black body and the Michelson experiments? Remind that in order for these clouds to be removed, it has taken the revision of physical notions about electromagnetic field as well as about space and time.

As is noted in [8], there is a simple physical mechanism of nuclear transformations at low energies whose existence follows from the quantum theory of electron as an open self-organizing system [9]. If two or the greater number of light nuclei appear inside free electron, more precisely, inside the area of basic localization of the particle, because of interaction of nuclei with electrically charged matter of electronic cloud, a force of attraction appears between the nuclei which may result in fusion of nucleus. This means that **cold nuclear reaction represents an intra-electronic process** whose character is defined by physical properties of the own field produced by electrically charged matter of electron. The purpose of this paper is more detailed consideration of the mechanism above stemming from the spatial extension of electron.

In section 2 physical ideas are formulated and basic results are schematically presented of quantum theory of electron as an open self-organizing system. The theory outlined is necessary to elucidate the origin of the mechanism resulting in the occurrence of nuclear reactions of fusion and fission at low energies. The essence of the approach developed consists in that **the own field created by electron is treated as a congenital, integral physical property of electron, intrinsically inherent in the particle by the very nature of things and for this reason the own field and self-action are included in the definition of the particle** at the initial stage of formulating the theory. As is seen from the results received, electron represents a quantum (elementary excitation) of the field of electrically charged matter. It is a soliton, whose physical and geometrical properties are described by the non-linear and non-local dynamical equation similar to the known Dirac equation.

In section 3 the application of quantum model of self-organizing electron to nuclear reactions at low energies is considered. It is noted that because of the presence of simple physical mechanism of nuclear reactions at low energies, which is of a universal character, nuclear reactors represent, in effect, nuclear delayed-action bombs which from time to time may blow up by virtue of the casual reasons. Hence, though nuclear stations may provide mankind with energy, however atomic engineering is a very dangerous way of energy production. The only acceptable way of solving the energetic problem consists in the use of nuclear reactions at low energies.

## 2. Quantum model of electron as an open self-organizing system

The basis for the standard formulation of quantum electrodynamics (QED) is the hypothesis that electron is a structureless point particle which does not experience self-action. This assumption results in serious difficulties – the divergences of mass and charge of electron and the impossibility to explain stability of the particle (see, for example, [10-12]).

The difficulties mentioned above are very serious. According to Dirac, **the difficulties of QED "in view of their fundamental character can be eliminated only by radical change of the foundations of the theory, probably, radical to the same extent as transition from the Bohr orbits theory to modern quantum mechanics"** ([13], p. 403). "Correct conclusion", Dirac emphasizes, "is that the basic equations are incorrect. They should be changed in such a way that divergences do not appear at all".

The main reason of occurrence of difficulties is the assumption that electron is a point-like particle. Therefore, abandonment of this hypothesis is inevitable. As an analysis of the problem shows, the key to constructing a consistent quantum theory of electromagnetism lies in taking account of the Coulomb self-action of electron, i.e. the back action of the own field created by charged particle in environmental space upon the same particle. In the special case that the particle is at rest in an inertial reference frame, own field of the particle turns into static Coulomb field.

One of the boldest ideas concerning the problem of electron was put forward by E.Schrödinger who suggested the historically first physical interpretation of quantum mechanics. According to Schrödinger's hypothesis, the quantity $e|\Psi(r)|^2$ ($e$ and $\Psi(r)$ are charge and wave function of electron, respectively) is the density of spatial distribution of electron's charge and, consequently, **the linear sizes of electron are the same as those of atom** [14,15]. However, they did not succeed in substantiating the interpretation and, for this reason, it was rejected by the majority of physicists [16].

An important step to correct understanding of the physical nature of electron was made by A. Barut and his collaborators [16-18] who formulated and developed quantum theory of electromagnetic processes on the basis of self-energy picture (the Self-Field QED). Using expression for the total own energy of electron, they managed to calculate the Lamb shift and other radiative corrections and to show that radiative phenomena may be described in terms of the action function, without using the second quantization method. As is pointed out by Barut [17], "the correct quantum equation of motion for radiating electron is not the Dirac or the Schrödinger equation for bare electron, but an equation containing an additional non-linear self-energy term".

New lines of approach to the problem of electron are offered in [9, 19-24]. The formulation of electrodynamics is considered which represents a synthesis of standard quantum electrodynamics and ideas of the theory of self-organization [25]. The physical mechanism of self-organization of electron consists in self-action. Taking into account the self-action means that electron is treated as a feedback system.

Let us outline schematically the results of the formulation of quantum electrodynamics in which electron is an open self-organizing system.

How can the self-action of electron be described? It is reasonable to suppose that the electric charge density of electron should be described by a continuous function of coordinates which assumes everywhere finite values and owing to this the own energy of the particle should be finite. One of the hints as to how to describe the self-action of electron and, hence, to remove divergence of self-energy can be obtained from Maxwell's equations for electromagnetic field. According to them, the potential electric field $E_\parallel$ created by an electrically charged particle and the total energy $W$ of this field can be written by

$$E_\parallel(r,t) = -\vec{\nabla} \int dr_1\, \rho_1 / |r - r_1|, \tag{1}$$

$$W = \frac{1}{2} \int dr_1 \int dr_2\, \rho_1 \rho_2 / |r_1 - r_2|, \tag{2}$$

where $\rho_n = \rho(r_n, t)$, $(n = 1, 2)$, $\rho(r, t)$ is the density of electric charge. Quantity $W$ represents the potential energy of self-action (the self-energy) of the particle. As is seen from (1) and (2),

$$E_\parallel = 0, \quad W = 0 \quad \text{at} \quad \rho = 0. \tag{3}$$

It follows from the last equalities that the Coulomb field $E_\parallel$ is not an independent degree of freedom of electromagnetic field: it is created by charged particle and cannot exist in its absence. Hence, the Coulomb field should be included in the definition of the particle. This idea can be realized on the basis of the action principle by including the self-action of electron in the Lagrangian function already in the zero-order approximation.

Obviously, when deriving the equation of motion for self-acting electron from the action principle, we should include the additional term, $-W$, in the Lagrangian function $L$ of the electron field, i.e.

$$L = L_0 - W, \tag{4}$$

where $L_0$ is the Lagrangian for free particle. Using as the function $L_0$ the Lagrangian function in nonrelativistic approximation,

$$L_0 = \int dr \left( \frac{i}{2} \Psi^* \overset{\leftrightarrow}{\partial_t} \Psi - \frac{1}{2m} \left( \vec{\nabla} \Psi^* \right)\left( \vec{\nabla} \Psi \right) \right) \tag{5}$$

($\Psi^* \overset{\leftrightarrow}{\partial_t} \Psi = \Psi^* \overset{\rightarrow}{\partial_t} \Psi - \Psi^* \overset{\leftarrow}{\partial_t} \Psi$, $\partial_t = \frac{\partial}{\partial t}$), and putting $\rho = e|\Psi|^2$, we arrive from the action principle at the following equation for the wave function $\Psi = \Psi(r,t)$ of nonrelativistic electron:

$$i\partial_t \Psi = \left( -\frac{1}{2m} \vec{\nabla}^2 + U \right) \Psi, \tag{6}$$

$$U(r,t) = e^2 \int dr_1 |r - r_1|^{-1} |\Psi(r_1, t)|^2 \equiv U. \tag{7}$$

An analysis shows, however, that equation (6) has no solutions satisfying necessary physical requirements. From the physical point of view, this is due to the fact that the Coulomb forces of repulsion acting inside electron tend to tear the particle to pieces. Formally, this is explained by the fact that potential energy $U$ (7) represents a potential hump rather than the potential well. For this reason equation (6) cannot have solutions describing stable states of electron.

Thus, the negative result is received: we have tried to take into account self-action of electron in a natural way by supplementing the Lagrangian function with the self-energy term, but we came to an equation that has no reasonable physical solutions at all. This result seems to mean that the standard theoretical scheme reaches here the limits of its applicability and so, remaining in its framework, it is impossible to solve the problem of electron and elucidate the physical nature of electromagnetic interaction.

Essentially new point which is introduced in [9] into quantum mechanics consists in the replacement of the model of isolated system described by harmonic oscillator with **the model of open system**. Let us advance the arguments indicating the inevitability of using the model of open system as a basis of the description of interaction between microparticles [26].

Note, first of all, that quantum particle theory based on the use of the models of isolated system is, strictly speaking, physically meaningless. Really, any observation conducted on a system represents a process of interaction of the system with the means of observation. But in case of microparticles (quantum particles) this interaction is not weak and consequently it is inadmissible to neglect it, i.e. microparticles should be necessarily considered as essentially non-isolated systems.

Starting point of the standard formulation of quantum mechanics is the physical idea that interaction between physical fields can be reduced to collision of the particles corresponding to these fields, the particles before and after collision being considered as free ones. According to these representations, quantum mechanics is based on the notions of "bare", non-interacting particles, with the interaction between them being considered as an additional factor which can only insignificantly alter the physical properties of non-interacting particles. However, such an approach to interaction between physical fields is obviously of an idealized character because particles constantly interact "with vacuum as with some kind of physical medium in which the particles move" [27]. Interaction of particles with vacuum fluctuations is not small and it cannot be removed.

It is well also to bear in mind that the necessary intermediary at studying micro-objects are the means of observations (the devices) with the classical field corresponding to them which should be taken into account in consistent quantum theory [28]. Inclusion in theoretical scheme of arbitrarily weak classical external field results in occurrence of non-zero width $\Gamma$ of energy levels of "dressed" particles. The basic impossibility to isolate a real particle from vacuum fluctuations of the field and from the classical sources connected to the means of observation is indicative, thus, of necessity to take into account the non-zero width of energy levels of real particles [26].

The use of the harmonic oscillator model, when describing the interaction of electromagnetic radiation with substance, seems to be the main source of serious difficulties of the standard formulation of quantum theory, as such an approach means apparent neglect of those physical processes which, proceeding constantly, are responsible for inseparable coupling of real physical system to surrounding medium. Introducing artificial notion about switching on and switching out of interaction of oscillator with radiation field, we are able to calculate within the framework of existing theory the width of energy levels of oscillator, but we cannot assert with certainty that such an approach results in correct description of interaction.

From the reasoning given above it is seen that it is the models with energy levels of non-zero width that should form the basis for the description of interaction of radiation with substance. It is necessary to formulate such a quantum theory which would take into account the energy levels of non-zero width $\Gamma$. The case in point is that one should introduce an infinitesimal damping $\Gamma$ into the initial set of equations describing interaction of charged particles with electromagnetic field. Such an approach means the violation in infinitesimal of homogeneity of physical system relative to translations in time. Necessity of violating the homogeneity of time follows from that fact that in the usual approach (with $\Gamma = 0$) the states of the system of interacting fields have degeneracy of infinitely large multiplicity in relation to time translations. According to the fundamental Bogoliubov's concept of quasi-averages [29], when describing the behaviour of degenerate systems, one should include into Hamiltonian an infinitesimal term removing degeneracy. In the theory presented here degeneracy of states of quantized fields relative translations in time is removed by introducing the infinitesimal damping $\Gamma$ into Lagrangian. Thereby the degeneracy under study is removed already in the initial, zero-order approximation, which is of fundamental importance for the approach based on perturbation theory.

Formulation of the physical idea that quantum friction arises at the very elementary level - at the level of one particle is given in monograph [26]. Impossibility to isolate real particle from the surrounding world is that property which should be taken into account already in the one-particle theory (for each kind of particles), even before switching on the interaction with other particles. Model of the particle as an open system ($\Gamma \neq 0$) is attractive owing to the fact that from the very beginning the degeneracy of states relative to time translations is absent in it, the degeneracy, which is removed in standard approach by taking into account the interaction of particle with vacuum field fluctuations and classical fields. The basis for the developed formulation is the fundamental concept of quasi-averages supplemented with the requirement that the equations of motion of the particle with $\Gamma \neq 0$ follow from the action principle. It should be emphasized that the non-zero damping $\Gamma$ is introduced into electrodynamics with the aim to establish the structure of the Lagrangian function which takes into account the property of openness of physical system. After establishing the structure, the limiting transition $\Gamma \to 0$ is fulfilled.

In our opinion, the development of quantum theory will be inevitably connected with the use of models of open system, as such models reflect more completely the physical essence of interrelations in the real world. It is necessary, thus, to define more exactly the concept of openness of physical system, which, on the one hand, would describe real system accurately enough and, on the other, would be simple enough to describe the particular physical processes.

As open system has the richer physical contents in comparison with isolated system, some essentially new mathematical ideas are needed for its description. First of all, it is necessary to increase the number of independent dynamical variables describing the particle as open system. In papers [9,19-24], as a basis for the description of self-acting electron, the simplest model of open system is used which can be described by the Morse-Feshbach-Bateman Lagrangian function [30,31] and which was successfully used for the description of dispersive media (the review of articles, in which applications of the model of open system to electrodynamics of dispersive media are considered, is given in monograph [26]). In this model the number of dynamical variables is doubled as compared with the isolated system, namely, to each dynamical variable of "bare" particle, $\Psi$, there correspond two dynamical variables, which are denoted by $\Psi$ and $\widetilde{\Psi}$. These quantities are considered as components of the wave function describing the quantum state of self-acting particle. One of them, say, $\Psi$, corresponds in a sense to the particle alone (to the "bare" particle) and the other, $\widetilde{\Psi}$, to the surrounding medium, in which the particle moves.

Dynamical variables $\Psi$ and $\widetilde{\Psi}$ are supposed to belong to the functional space with indefinite metric defined by quadratic form

$$\widetilde{\Psi}^*\Psi + \Psi^*\widetilde{\Psi} \quad . \tag{8}$$

This quantity is used instead of the positively defined quadratic form

$$\Psi^*\Psi , \tag{9}$$

underlying the conventional formulation of quantum mechanics.

At first sight, the Lagrangian function $L = L(\Psi, \widetilde{\Psi})$ of real electron, as a system composed of two subsystems interacting with each other, should be constructed in the standard manner:

$$L(\Psi,\widetilde{\Psi}) = L_1(\Psi) + L_2(\widetilde{\Psi}) + L_{\text{int}}(\Psi,\widetilde{\Psi}), \tag{10}$$

where $L_1(\Psi)$ is the Lagrangian function of "bare" electron (i.e. of the particle isolated from the medium), $L_2(\widetilde{\Psi})$ is the Lagrangian function of the medium, in which the particle moves, and $L_{\text{int}}(\Psi,\widetilde{\Psi})$ is the Lagrangian function describing interaction of "bare" electron with surrounding medium, with the equalities

$$L_{\text{int}}(0,\widetilde{\Psi}) = L_{\text{int}}(\Psi,0) = 0 ,$$

being fulfilled. If we now neglect the dynamical variables of the medium, that is, if we put $\widetilde{\Psi} = 0$, we shall come to the Lagrangian function of "bare" particle,

$$L(\Psi, 0) = L_1(\Psi) ,$$

being considered as the zero-order approximation for real particle. We should take into account, however, that **real, observable electron is indissolubly bound to environment; its interaction with environment cannot be weak as it is a condition for existence of particle as a physical system**: having isolated the particle from environment, we shall receive an unobservable object, whose physical properties cannot be determined [26, 28]. From this point of view the use of the Lagrangian function (10) as a basis of the theory is unsatisfactory. At the same time, replacement of quadratic form (9) by quadratic form (8) results in a natural way in such an expression $\widetilde{L}(\Psi,\widetilde{\Psi})$ for the Lagrangian function of real electron, which takes into account the inseparability of the particle from environment. In particular, the equalities

$$\widetilde{L}(\Psi,0) = \widetilde{L}(0,\widetilde{\Psi}) = 0$$

are fulfilled, which mean that the "bare" electron approximation has no physical sense; in either case, with no electron or with no medium, we have no physical system. Thus, the last equalities may be considered as **the condition for openness of the system** under study.

All physical quantities are expressed in terms of the quadratic form (8); in particular, the electric charge density of electron in nonrelativistic approximation is given by

$$\rho(\mathbf{r},t) = e\,(\widetilde{\Psi}^*(\mathbf{r},t)\Psi(\mathbf{r},t)+c.c) \ . \tag{11}$$

Including the Coulomb field produced by electron in the definition of the particle, we use, as the Lagrangian function of self-acting electron field, $L$, the expression (4) where

$$L_0 = \int d\mathbf{r}\ \left\{\frac{i}{2}\left(\widetilde{\Psi}^*\overleftrightarrow{\partial}_t\Psi + \Psi^*\overleftrightarrow{\partial}_t\widetilde{\Psi}\right) - \frac{1}{2m}\left[\left(\vec{\nabla}\widetilde{\Psi}^*\right)\left(\vec{\nabla}\Psi\right)+\left(\vec{\nabla}\Psi^*\right)\left(\vec{\nabla}\widetilde{\Psi}\right)\right]\right\} \tag{12}$$

is the Lagrange function of the particle without self-action, $W$ is the own energy of electron defined by expression (2), in which $\rho_n = \rho(\mathbf{r}_n,t)$, $(n=1,2)$, $\rho(\mathbf{r},t)$ is the charge density (11). The action principle with Lagrangian function (4) results in the following non-linear equation of motion of electron in nonrelativistic approximation:

$$\left(i\partial_t + \frac{\vec{\nabla}^2}{2m} - U(x)\right)\begin{pmatrix}\Psi(x)\\\widetilde{\Psi}(x)\end{pmatrix}=0, \quad x=(\mathbf{r},t), \tag{13}$$

$$U(x) = e\int d\mathbf{r}_1\,|\mathbf{r}-\mathbf{r}_1|^{-1}\,\rho(\mathbf{r}_1,t). \tag{14}$$

As an analysis shows, equation (13) has solutions describing stationary states of particle at $N=-1$, where

$$N = \int d\mathbf{r}\ (\widetilde{\Psi}^*\Psi + \Psi^*\widetilde{\Psi}) \tag{15}$$

is the normalization constant. The occurrence of two components of wave function without regard for spin means that the particle has an additional degree of freedom. In the theory developed this degree of freedom is described by the sign of normalization constant $N$ ($N=\pm 1$), which plays the role of a quantum number taking into account the electron self-action.

As is seen from (2), (4), (11) and (12), the Lagrangian function $L$ of electron vanishes both for $\widetilde{\Psi}=0, \Psi\neq 0$ and for $\Psi=0, \widetilde{\Psi}\neq 0$. Physically, this means that the "bare" electron and the "bare" medium generated by electron do not exist in nature in separation from each other. Electron, isolated from its own Coulomb field, and the Coulomb field separated from electron, taken separately, have no physical sense.

Equation (13) can be easily generalized to relativistic case. Omitting details of calculations, which can be found in [9,22-24], we shall present the equation of motion in the final form:

$$\left(i\hat{\partial} - e\hat{A}(x) - m\right)\begin{pmatrix}\Psi(x)\\\widetilde{\Psi}(x)\end{pmatrix} = 0, \tag{16}$$

where $\hat{\partial} = \gamma_0\dfrac{\partial}{\partial t}+\vec{\gamma}\dfrac{\partial}{\partial \mathbf{r}}$, $\hat{A}(x) = \gamma_0 A_0(x) - \vec{\gamma}\vec{A}(x)$, $\gamma_0$ and $\vec{\gamma}$ are Dirac's matrices, $\Psi(x)$ and $\widetilde{\Psi}(x)$ are Dirac's bispinors,

$$A(x) = A_\parallel(x) + A_\perp(x),$$

$$A_\parallel(x) = \int d^4x_1\,\delta\!\left((x-x_1)^2\right) j_\parallel(x_1), \quad A_\perp(x) = (0, \mathbf{A}_\perp(x)),$$

$$\mathbf{A}_\perp(x) = (4\pi)^{-1}\int d^4x_1\,\delta\!\left((x-x_1)^2\right)\!\left(\vec{\nabla}_{r_1}\!\times\mathbf{B}(x_1) - \partial_{t_1}\mathbf{E}_\perp(x_1)\right),$$

$$j_\parallel(x) = (\rho(x), \mathbf{j}_\parallel(x)),$$

$j^\mu(x) = e\left(\overline{\widetilde{\Psi}}(x)\gamma^\mu \Psi(x) + \overline{\Psi}(x)\gamma^\mu \widetilde{\Psi}(x)\right) \equiv (\rho(x), \boldsymbol{j}(x))$ is the 4-vector of current density.

Here $\boldsymbol{B}$ and $\boldsymbol{E}$ are the magnetic and electric fields, $\overline{\Psi} = \Psi^+ \gamma_0$, $\overline{\widetilde{\Psi}} = \widetilde{\Psi}^+ \gamma_0$, the potential $\boldsymbol{j}_\|$ and vortex $\boldsymbol{j}_\perp$ components of the vector $\boldsymbol{j} = \boldsymbol{j}(\boldsymbol{r},t)$ are defined by the following equalities:

$$\boldsymbol{j}_\|(\boldsymbol{r},t) = -\mathrm{grad}\ \mathrm{div}\ \int d\boldsymbol{r}'\ (4\pi)^{-1} \boldsymbol{j}(\boldsymbol{r}',t)/|\boldsymbol{r}-\boldsymbol{r}'|,$$

$$\boldsymbol{j}_\perp(\boldsymbol{r},t) = \mathrm{curl}\ \mathrm{curl}\ \int d\boldsymbol{r}'\ (4\pi)^{-1} \boldsymbol{j}(\boldsymbol{r}',t)/|\boldsymbol{r}-\boldsymbol{r}'|.$$

Note that electromagnetic field, as well as electron field, should be considered on the basis of the model of open system. In the simplest variant of the theory, each dynamical variable of electromagnetic field should be split into two independent components, one of which describes an electromagnetic wave, and the other – the environment, in which the wave is propagated and with which it interacts. Such an approach is stated in [9]. For simplicity, in the above are given the formulas relating to the case that electromagnetic field is described in the usual way, without doubling the dynamical variables.

Equation (16) describes one self-acting particle interacting with vortex electromagnetic field. Generalization of this equation to the set of arbitrary number $n$ of self-acting electrically charged particles interacting with each other is given in [9].

Equation (16) coincides in its appearance with the usual Dirac equation for charged particle in an external field described by 4-potential $A$. However, in reality, it differs essentially from Dirac's equation. The distinction consists in that **equation (16) is non-linear and non-local**, with the non-locality being of both spatial and time character. Potential ($A_\|$) and vortex ($A_\perp$) components of the 4-potential, entering equation (16), differ from each other by their physical nature: the former describes the Coulomb field and is expressed quadratically in terms of the wave function components of electron, and the latter describes transverse electromagnetic waves and is expressed in terms of vortex electromagnetic field. As a detailed analysis shows, solutions to the basic dynamical equation describe the clots of self-acting electrically charged matter, localized in space, i.e. the particle is a soliton.

The internal energy spectrum of electron is discrete with an indefinitely large number of levels, and to each value of internal energy $E_k$ ($k$ is the set of quantum numbers) there correspond certain linear dimensions and geometrical form of the region of localization of electron's charge. Dimensions and the number of extrema of wave function increase with increasing the value of energy $E_k$. The distribution of electric charge of atomic electron in the ground state consists of the range of basic localization with the linear dimensions of the order of Bohr radius $a_0$ ($a_0 \sim 10^{-10} m$) and of the tail stretching up to infinity. It is essential that because of non-linearity of the dynamical equation of electron, wave function does not obey the superposition principle. By virtue of this, electron acquires the properties of absolutely rigid body: the perturbation acting on electron at an instant of time $t$ in the range of basic localization becomes known at the next instant $t+0$ at any distance from the particle.

In Fig. 1 are represented schematically the results of calculation, carried out on the basis of equation (13), of the distribution of electric charge in atomic and free electrons in the ground (a) and first excited (b) states.

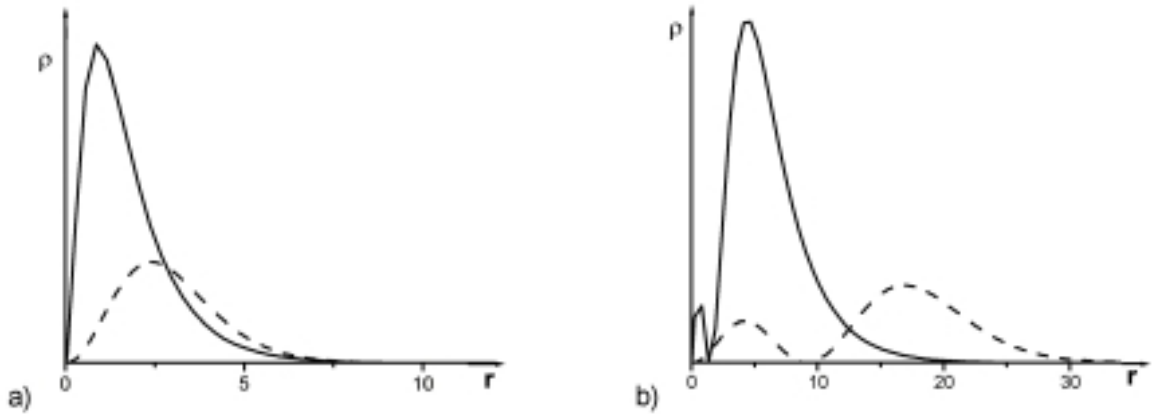

Fig. 1. Density of electric charge ($\rho$) of electron in the ground state (a) and in the first excited state (b): the continuous lines correspond to electron in the hydrogen atom, and the dotted ones to free electron, $r$ is the distance from the center of mass of electron measured in Bohr radii.

According to [9,19], the atom represents a system of nuclear and electronic solitons interacting with each other, the internal energy spectrum of the hydrogen atom, due to electromagnetic interaction, being of a zoned character. The occurrence of zoned structure of energy spectrum of hydrogen atom is explained as follows. Free nucleus, because of existence of Coulomb self-action, has a discrete internal energy spectrum. As the interaction of nucleus with electron is small in comparison with the energy of Coulomb self-action of the nucleus, it can be taken into account by perturbation theory. From here it follows at once that each energy level of free nucleus is split in a zone. There are indefinitely many zones (Balmer's replicas) and in each of them there are indefinitely many energy levels. The lowest zone coincides with the usual Balmer spectrum.

### 3. Physical mechanism of nuclear reactions at low energies

The quantum theory presented above schematically of electron as an open self-organizing system is indicative of the existence of the following mechanism of nuclear reactions at low energies [8].

If there occur in the region of basic localization of free electron, whose linear sizes in the ground state of the particle are several times as large as those for hydrogen atom (see Fig. 1), two or the greater number of nuclei, each of them attracts on itself the adjoining areas of electronic cloud, resulting in compression of the electronic cloud as a whole. As a result, there appears automatically an attraction of the nuclei, which proved to be "inside" electron, on each other (see Fig. 2).

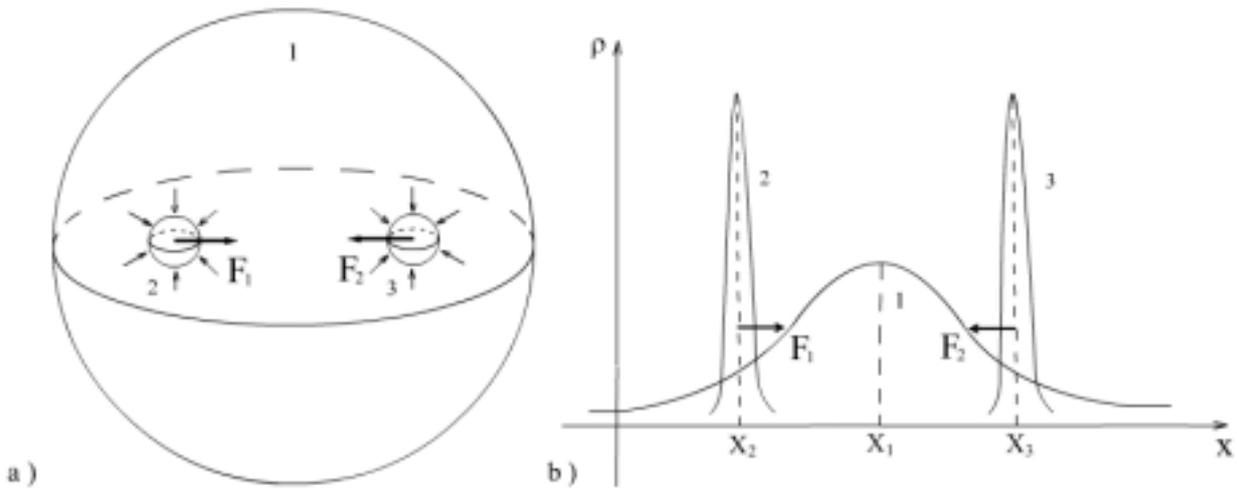

Fig. 2. The schematic image of interaction of nuclei with electronic cloud: (a) 1 is the region of basic localization of electron, 2 and 3 are nuclei, $\mathbf{F}_1$ and $\mathbf{F}_2$ are the attractive forces between nuclei, which appear at the expense of electronic cloud compression induced by Coulomb forces; (b) $\rho$ is the charge density, 1 is electronic soliton, 2 and 3 are nuclear solitons, $X_n$ ($n = 1, 2, 3$) are coordinates of the centers of mass of particles.

Calculation shows that the Coulomb barrier around nuclei is deformed, its height decreases and the probability of penetration through the barrier accordingly increases due to tunnel transition. Under certain conditions this process may result in fusion of nuclei. Obviously, the process in question can occur only at small energies of translational motion of the centers of mass of electron and nuclei: nuclei should be "inside" electron long enough for them to have time to come nearer to each other as a result of electron-nuclear interaction. This mechanism of nuclear fusion is of a universal character. In order for it to be realized, it is necessary to have only a stream of free electrons intensive enough, i.e. heavy electric current, and as long as sufficiently great number of free nuclei.

If heavy nuclei appear "inside" free electron, owing to their interaction with the electronic cloud there occurs polarization of nuclei. Because the own field of electron interacts with protons more strongly than with neutrons, nuclei are deformed (become extended), and this process may result in the decomposition of nuclei to fragments (in nuclear fission).

As is noted in [7], the official version of the reasons for Chernobyl accident contains serious contradictions, a number of facts concerning the accident has no convincing explanations, and this circumstance forces to search for the true reasons for the happening, since "not having understood the mechanism of the one tragedy, we sooner or later shall become witnesses of the other". The authors hypothesize that the reason of the accident was penetration into the nuclear reactor of magnetic monopoles, which have caused the decay of nuclei $^{238}U$, and this has resulted in production of delayed neutrons, growth of power output of the reactor and explosion. As an argument in favour of the assumption, the fact is presented that nuclei $^{238}U$ are disintegrated under the action of "strange" radiation appearing at explosion of foil.

In the opinion of the authors of [5,7], "strange" radiation is created by those magnetic monopoles which form bound states with nuclei of atoms. These compound particles give the abnormally wide tracks similar to those of a creeping caterpillar, and also the tracks of complicated shape reminiscent of spirals and gratings. Character of tracks changes when imposing magnetic field, which, as the authors believe, is an argument in favour of the assumption above. There are also some special tracks very similar to scratches and ink spots. "Strange" radiation is of spherical form, it resembles a ball lightning, and its duration is more than ten times as great as that of the current pulse arising at electric discharge. With the course of time the luminous sphere (the ball-like plasma formation) is dividing into many small "balls".

It is our opinion that "strange" radiation is caused by free electrons in excited state arising in the area of electric discharge. According to [9, 19], linear sizes of the region of basic localization of such electrons can make many tens of sizes of atom. The heavy nucleus, for example, the nucleus $^{238}$U, appearing inside the electronic cloud, is inevitably deformed because of interaction of protons with adjoining layers in the distribution of electric charge of electron, and this deformation can cause nuclear fission. If two or the greater number of light nuclei appear "inside" electron, then attractive forces arise between nuclei which may result in fusion reaction. When electric discharge is strong enough, the areas of basic localization of some electrons can overlap, and if a nucleus lands in the area of overlap, because of Coulomb attraction of nucleus on the adjoining layers of electronic clouds, a bound state may be formed, of two electrons and the nucleus, characterized by the relative stability and significant spatial extension.

Obviously, if the concentration of free electrons is great enough, there may be formed some relatively stable bunch of plasma consisting of great number of free electrons and nuclei, which in virtue of chaotic movement of nuclei and because of the absence of preferred directions should have approximately spherical form. Note that to "strange" radiation can contribute atomic electrons, belonging to additional energy zones of atom (Balmer's replicas associated with nuclear self-action, see Section 2).

As is seen from the above, to account for the reasons for Chernobyl accident, there is no need to involve magnetic monopoles. The scenario of development of events during the accident, described in [7], seems to be quite plausible if only to understand by initiators of nuclear fission not hypothetical monopoles but free electrons, whose powerful pulse might arise as a result of electric discharge in the region of turbo-generators.

The existence of simple physical mechanism of nuclear reactions at low energies, indicated in this paper, implies that nuclear reactors are, in effect, nuclear delayed-action bombs which will blow up from time to time. Explosion of nuclear reactor may take place because of casual short circuit at an electric subcircuit, owing to which there appears an intensive stream of free electrons. This stream, having got for any reasons in nuclear reactor, may initiate explosion of the reactor. It follows from here that though nuclear stations may provide mankind with cheep energy, atomic energetics represents a very dangerous way of producing energy (as well as the energetics using controlled thermonuclear fusion). The only acceptable way of resolving the energetic problem consists in the use of nuclear reactions at low energies.

According to the results obtained, nuclear reactions at low temperatures occur "inside" electron under the action of own field of particle. Hence, to elucidate physical mechanism of CF, it is necessary to study in detail intra-electronic processes and physical properties of own fields of particles. Note that the own field, by its physical properties, essentially differs from the field of electromagnetic waves: this is the field of standing waves of matter, it is of purely classical character and may not be reduced to the set of photons. The own field of charged particle plays in nature a special role, consisting in that it transforms environmental space into the physical environment (physical vacuum) with the properties of absolutely rigid body [32].

As it was repeatedly noted in the literature [1,2], experiments on CF are badly reproduced, and this fact gives rise to doubt the very existence of the phenomenon. Bad reproducibility of results seems to be explained by the fact that CF depends upon great number of parameters: upon electric current density, concentration of free nucleus, concentration of impurities and dislocations in samples, sizes of samples etc. In order to obtain reproducibility of results, it is necessary that all these parameters, describing the environment in which nuclear reactions occur, be the same in various experiments, but to achieve this is a difficult task.

In conclusion we shall dwell upon the problem of linear dimensions of electron, which is of special interest in connection with the mechanism of nuclear reactions indicated here. The inference that the dimensions of electron in the ground state of atom are of the order of Bohr radius, i.e. of the order of atomic dimensions, following from dimension considerations [9,19] and confirmed by quantum model of electron, seems completely unexpected. At first sight, it is in conflict with both the theory of quarks and experimental data on scattering of electrons. According to quark models, the radius of electron corresponding to its quark structure makes up the quantity of the order of $10^{-22}$ m [33]. It is necessary to emphasize, however, that the above mentioned magnitude of linear dimensions of electron refers to the internal structure induced by Coulomb field. The last is long-distance and consequently the linear dimensions of internal structures produced by it (i.e. spatial inhomogeneities in the distribution of electric charge in various quantum states) should considerably exceed the dimensions of quark structures connected with electron. There seems to exist a hierarchy of internal structures of particle produced by Coulomb forces, nuclear forces, inter-quark interactions etc. characterized by the smaller and smaller linear sizes.

As to the experiments on scattering of high energy electrons, according to which the internal structure of electron is not manifested up to distances of the order of $10^{-16} \div 10^{-17}$ m, two arguments, at least, can be adduced in favour of that there is no contradiction here with experiment. Firstly, in experiments on scattering, investigators were trying to register the details of internal structure of electron within intervals much smaller than Bohr radius, which is why it is not surprising that results of experiments proved to be negative: at high energies electrons behave like point particles, their internal structure has no time to be manifested. Secondly, the results of experiments were analyzed from the point of view of standard representations about electron, which refer to a point particle, but are obviously inapplicable to real, self-acting electron. According to the predictions of quantum theory of electron as an open self-organizing system, real electron is a special object - soliton, i.e. such a cloud of electrically charged substance which, when interacting with other particles, tends to keep its sizes and geometrical form. At present there is as yet no scattering theory of this kind of particles and for this reason it is impossible to predict with certainty how can the internal structure of electron be manifested in experiments on scattering.